\documentclass[conference]{IEEEtran}
\usepackage{epsfig}
\usepackage{url}
\usepackage{tabularx}
\usepackage{listings}
\usepackage{fixme}
\usepackage{graphicx}
\usepackage{subcaption}
\usepackage[frozencache]{minted}
\fxsetup{
    status=draft,
    author=,
    layout=inline,
    theme=color
}

\usepackage{tikz}

\newcommand*\smallcircled[1]{\tikz[baseline=(char.base)]{
            \node[shape=circle,draw,inner sep=1pt] (char) {#1};}}

\IEEEoverridecommandlockouts

\begin{document}

\date{}

\title{\Large \bf Chaff Bugs: Deterring Attackers by Making Software Buggier}

\author{{\rm Zhenghao Hu}\thanks{Portions of this work were carried out while Zhenghao Hu was attending Shanghai Jiao Tong University.} \\
New York University \\
huzh@nyu.edu
\and
{\rm Yu Hu} \\
New York University \\
yh570@nyu.edu
\and
{\rm Brendan Dolan-Gavitt} \\
New York University \\
brendandg@nyu.edu}

\maketitle


\begin{abstract}

Sophisticated attackers find bugs in software, evaluate their exploitability, and then create and launch exploits for bugs found to be exploitable. Most efforts to secure software attempt either to eliminate bugs or to add mitigations that make exploitation more difficult. In this paper, we introduce a new defensive technique called \emph{chaff bugs}, which instead target the bug discovery and exploit creation stages of this process. Rather than eliminating bugs, we instead \emph{add} large numbers of bugs that are provably (but not obviously) non-exploitable. Attackers who attempt to find and exploit bugs in software will, with high probability, find an intentionally placed non-exploitable bug and waste precious resources in trying to build a working exploit. We develop two strategies for ensuring non-exploitability and use them to automatically add thousands of non-exploitable bugs to real-world software such as \texttt{nginx} and \texttt{libFLAC}; we show that the functionality of the software is not harmed and demonstrate that our bugs look exploitable to current triage tools. We believe that chaff bugs can serve as an effective deterrent against both human attackers and automated Cyber Reasoning Systems (CRSes).

\end{abstract}

\section{Introduction}

Bugs in software are both common and costly. Particularly in languages that lack memory safety guarantees, such as C and C++, it is all too easy for programmer errors to result in memory corruption. Worse, these bugs can often be exploited by attackers to achieve arbitrary code execution. 

However, not all bugs are created equal. Depending on the exact nature of the bug and the runtime environment when it is triggered, a bug may not lead to a violation of any of the program's security goals. In these cases we say that the bug is \emph{non-exploitable}. For example, although null pointer dereferences are certainly violations of memory safety and therefore bugs, on systems that prohibit mapping the first page of memory these bugs can only crash the program.

\begin{figure}
\centering
\includegraphics[width=3in]{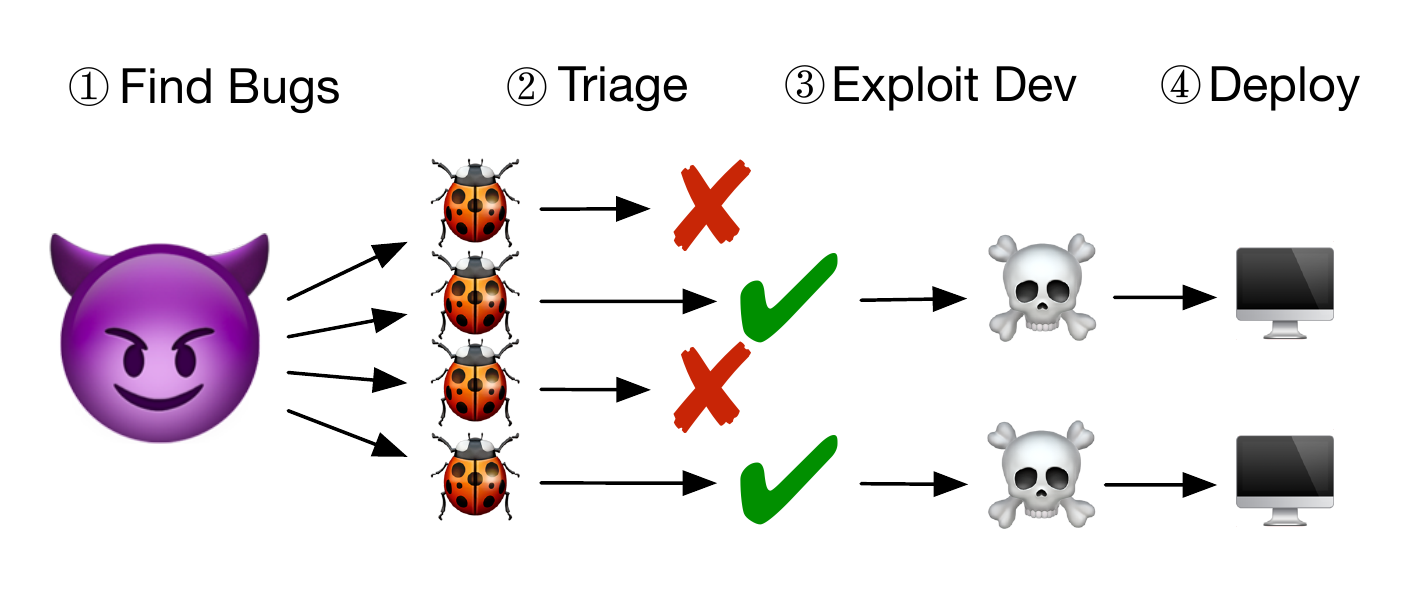}
\caption{Simplified attacker workflow. Attackers find bugs, triage them to determine exploitability, then develop exploits and deploy them to their targets.}
\label{fig:attackflow}
\end{figure}

If an attacker wants to exploit a flaw in a program, they must therefore triage any bugs they find to determine their exploitability and then work to develop an exploit for any deemed exploitable; this attacker workflow is depicted in Figure~\ref{fig:attackflow}. The process of going from an initial bug to a working exploit is generally long and difficult. And although some research has sought to automate exploit generation~\cite{Avgerinos:2014,CGC}, it is still a largely manual process.

On the defensive side, current efforts to secure software generally focus on one of two areas: bug finding and exploit mitigation. In terms of the attacker workflow in Figure~\ref{fig:attackflow}, the former attempts to reduce the supply of bugs available to the attacker in step \smallcircled{1}, while the latter attempts to reduce the effectiveness of the exploit launched in step \smallcircled{4}. Unfortunately, despite decades of effort, automated techniques to find bugs cannot guarantee their absence for any but the smallest programs. Exploit mitigations such as ASLR and CFI~\cite{cfi} raise the bar for attackers, but typically come with performance penalties and do not deter the most sophisticated attackers.

In this paper, we introduce a new defensive technique that is aimed at disrupting steps \smallcircled{2} and \smallcircled{3}: the bug triage and exploit development process. Rather than eliminating bugs, we propose instead to \emph{increase} the number of bugs in the program by injecting large numbers of \emph{chaff bugs}\footnote{Named after the strips of foil dispensed by military aircraft to confuse enemy radar.} that can be triggered by attacker-controlled input. By carefully constraining the conditions under which these bugs manifest and the effects they have on the program, we can ensure that chaff bugs are non-exploitable and will only, at worst, crash the program. As long as it is difficult for attackers to determine that a chaff bug is non-exploitable, we can greatly increase the amount of effort required to obtain a working exploit by simply adding so many non-exploitable bugs that any bug found by an attacker is overwhelmingly likely to be non-exploitable.

Although in some cases bugs that cause crashes constitute denial of service and
should therefore be considered exploitable, there are large classes of software
for which crashes on malicious inputs do not affect the overall reliability of
the service and will never be seen by honest users. Such programs include most
microservices~\cite{dragoni2016microservice} (which are designed to gracefully
handle the failure of a server by restarting it), server-side conversion
utilities, or even web servers such as \verb+nginx+ that maintain a pool of
server processes and restart any process that crashes. In these cases, failures
caused by attempts to exploit a non-exploitable bug are invisible to normal
users with benign inputs.

Prior work~\cite{Dolan-Gavitt:2016,Pewny:2016} has developed techniques for injecting potentially exploitable bugs into programs, but it is not immediately obvious that creating non-exploitable bugs is possible without making it easy for attackers to triage them. We devise two strategies for guaranteeing that an injected bug is non-obviously non-exploitable. The first strategy \emph{overconstrains} the value used to corrupt memory so that can only take on safe values, while the second ensures that the data to be overwritten is \emph{unused} by the program. We build a working prototype of a chaff bug injector that can create non-exploitable stack- and heap-based overflows, and use it to put thousands of bugs into three large, real-world programs, demonstrating the feasibility of our techniques.

\section{Scope and Threat Model}
\label{sec:scope}

Typical information security goals are to protect the confidentiality, integrity, and availability of data or services. For a given program, we assume that the attacker's goal is to compromise one of these properties by finding and exploiting \emph{memory safety bugs} in the program. We make no assumptions about how the attacker finds the bugs: they may use manual analysis, automated techniques based on static or dynamic analysis, etc.

We assume that the attacker has access to a compiled (binary) version of the program but not its source code. Although this is a severe limitation (it means that our technique cannot be used to protect open source software), it is a necessary one. Developers are unlikely to be willing to work with source code that has had extra bugs added to it, and more importantly future changes to the code may cause previously non-exploitable bugs to become exploitable. Hence we see our system as useful primarily as an extra stage in the build process, adding non-exploitable bugs.

Once an attacker finds a memory safety bug, he must \emph{triage} it to determine its exploitability. In general, this is thought to be a difficult task (though not one that has received much academic attention; we discuss some existing approaches to determining exploitability in Section~\ref{sec:future}). Aside from straightforwardly determining exploitability, the fact that our bugs are injected automatically offers the attacker another strategy: he can attempt to \emph{distinguish} an injected bug from a naturally occurring bug, by identifying some artifact of injected bugs that is not present in natural bugs. This would allow the attacker to bypass the triage process and ignore any injected bugs.

Some of the bugs we create (specifically the \emph{overconstrained} bugs described in Section~\ref{sssec:overconstrained}) can cause the program to crash. This would appear to break the requirement of availability for our threat model. In order to satisfy our threat model, we must restrict our scope further: \emph{we can only add chaff bugs to programs where crashes on malicious inputs do not degrade the program's availability on honest inputs}. Despite this limitation, we note that many applications fit this model. Modern web browsers split each tab into its own process~\cite{chrome_multiprocess}, and many web servers such as \texttt{nginx} (used in our evaluation) create a process pool to handle incoming connections; in these cases crashing on a malicious input does not affect the usability of the program as a whole. Likewise, command line utilities that process a single file at a time fit our model; while they may crash on a malicious input, this does not affect their ability to process other inputs.

For our current work, we explicitly choose not to address the \emph{distinguishability} problem mentioned above, even though it is within our threat model. Although we believe it is possible to make our bugs indistinguishable from naturally occurring bugs, we do not currently have a good understanding of what makes a naturally occurring bug look ``realistic'', and without such an understanding our attempts would necessarily be incomplete. We also see the distinguishability problem as orthogonal to the problem of making bugs non-exploitable, and choose to focus on the latter in this paper. Nonetheless, we recognize that distinguishing attacks are a valid threat, and we hope to address them in future work (we offer a few ideas on this front in Section~\ref{sec:future}).

\section{Background: Automated Vulnerability Addition}
\label{sec:background}

In order to automatically add large numbers of non-exploitable bugs to software to software, we first need to be able to add large numbers of bugs of any sort. In prior work, Dolan-Gavitt et al.~\cite{Dolan-Gavitt:2016} developed LAVA, a system for automatically adding potentially exploitable memory safety errors into C/C++ programs. Although LAVA is intended to create high-quality bug corpora for testing bug-finding tools, in this work we repurpose and extend it to create chaff bugs. In this section we briefly provide background information on how LAVA adds bugs to programs before moving on to describe our extensions to ensure non-exploitability.

LAVA takes a program and a set of input files to that program. It then runs the program on each input and performs a \emph{dynamic taint analysis}~\cite{taint} to determine how each byte of the input file is used in the program. It looks for points in the program where the input is \emph{dead} (not used to decide any branches up to that point in the program), \emph{uncomplicated} (not transformed), and \emph{available} in some program variable. These variables, which are called \emph{DUAs}, represent attacker-controlled inputs that can be used to later trigger memory corruption in the program. When a DUA is encountered, its value is copied to a global variable so that it can later be retrieved at the point where the memory corruption is triggered.

To actually trigger the bug, LAVA also makes note of \emph{attack points} along the trace---that is, places where the stored data from the DUA could be used to corrupt memory. In its original implementation, LAVA considers any pointer argument to a function call to be an attack point; in our work we instead create new attack points that cause controlled stack and heap overflows at arbitrary points in the program. At the attack point, a \emph{trigger condition} is also checked before the DUA is allowed to corrupt memory; this ensures that the bug is triggered under controlled conditions.

For the purposes of LAVA, a bug is simply the combination of a DUA and an attack point that occurs after the DUA in the trace. Each bug created by LAVA comes with a triggering input, which makes it easy to verify that the bug is actually present in the program. Dolan-Gavitt et al.~\cite{Dolan-Gavitt:2016} demonstrated that thousands of bugs can be added to a program using this technique, which makes it a good fit for creating chaff bugs, since we must ensure that any bugs found in the program are likely to be our own non-exploitable bugs.

The bugs themselves are injected by performing a source-to-source transformation on the program. At the DUA site, code is inserted to copy the DUA's value into a global variable via a call to \texttt{lava\_set}. At the attack point, \texttt{lava\_get} retrieves the value, compares it to the trigger value, and, if the condition is met, adds the DUA to the attacked pointer in order to cause a memory corruption. An example of this type of bug is shown in Figure~\ref{fig:lavabug}.

\begin{figure}
\begin{minted}{c}
void get_external_input(int fd) {
  ...
  read(fd, buf, 1024);
  lava_set(*(int *)buf);
  ...
}
...
int foo(int y) {
  int x[10];
  // Original:
  // x[0] = 10;
  // BUG:
  x[0 + (lava_get()==0x6c617661)*lava_get()] = 10;
}
\end{minted}
\caption{A bug inserted by the original LAVA system. If the input contains the correct trigger value, then an out-of-bounds write will occur through the array \texttt{x}.}
  \label{fig:lavabug}
\end{figure}

Following the publication of LAVA, Pewny and Holz described an alternative approach to bug injection called EvilCoder~\cite{Pewny:2016}. Rather than creating new bugs in the target program, EvilCoder identifies potentially vulnerable code (i.e., points where attacker-controlled data reaches a sensitive operation such as a \texttt{memcpy}), and then removes any safety checks that currently prevent the attacker-controlled data from causing a vulnerability. Although in principle this technique could be adapted to create chaff bugs, the number of bugs it can create is inherently limited by the number of points in the program where attacker-controlled data reaches a sensitive sink. This prevents us from using EvilCoder for our chaff bugs, since we need to guarantee that we can add large amounts of chaff to deter attackers.

\section{Design}
\label{sec:design}

To achieve the goal of increasing adversarial workload without compromising the security of the software, we must ensure that chaff bugs are 1) non-exploitable, and 2) difficult to triage to determine exploitability. In this section, we discuss our strategies for achieving these goals for stack- and heap-based buffer overflows.

As mentioned in Section~\ref{sec:scope}, guaranteeing non-exploitability requires that we have some control over the compiler and runtime environment. Specifically, we need to control the layout of local variables on the stack and the heap allocator used, and we need to be able to predict what memory ranges will be unmapped at runtime. We believe these constraints are reasonable for commercial software since the software developers have full control over their development toolchain.

\begin{figure}
	\centering
	\includegraphics[width=0.5\textwidth]{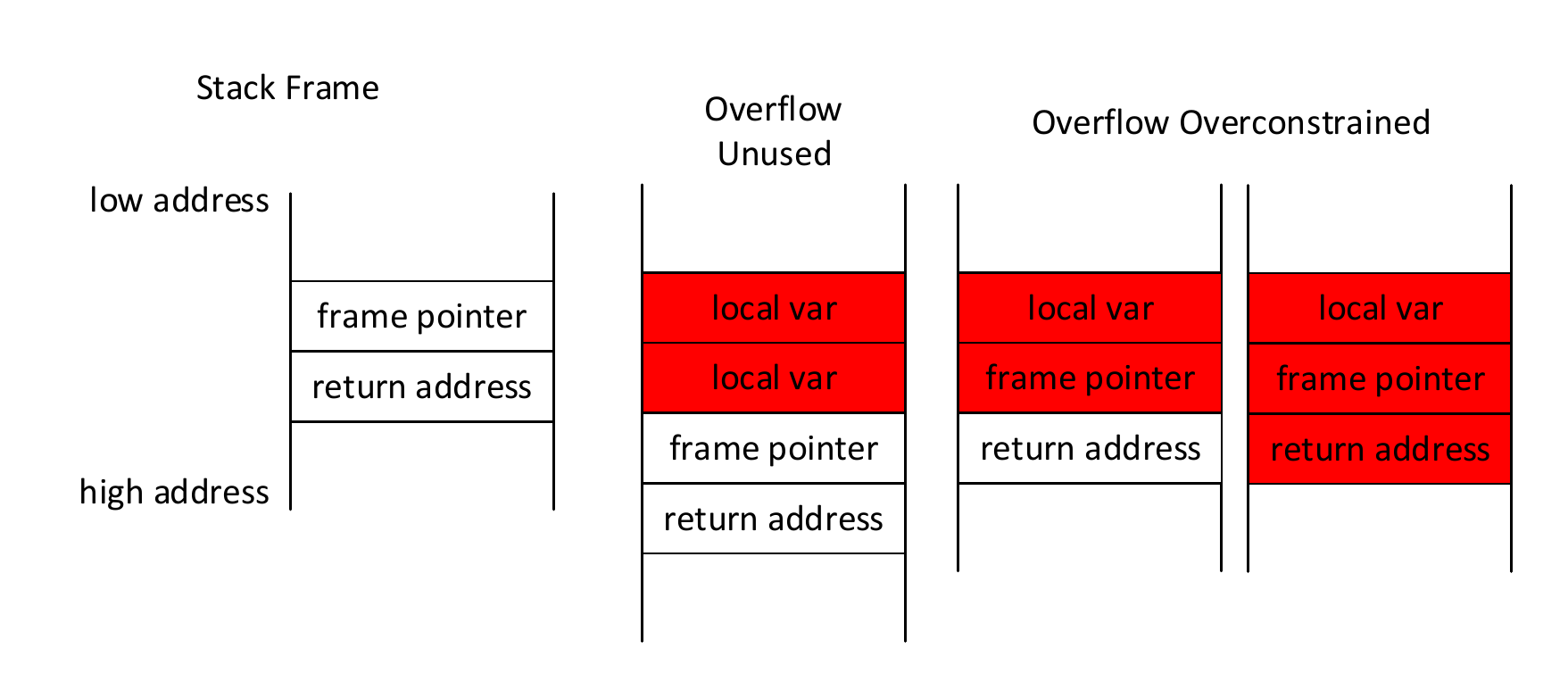}
	\caption{Stack Overflow. The areas overwritten when our bugs are triggered are indicated in red.}
	\label{fig:stackbugs}
\end{figure}

\begin{figure}
	\centering
	\includegraphics[width=0.5\textwidth]{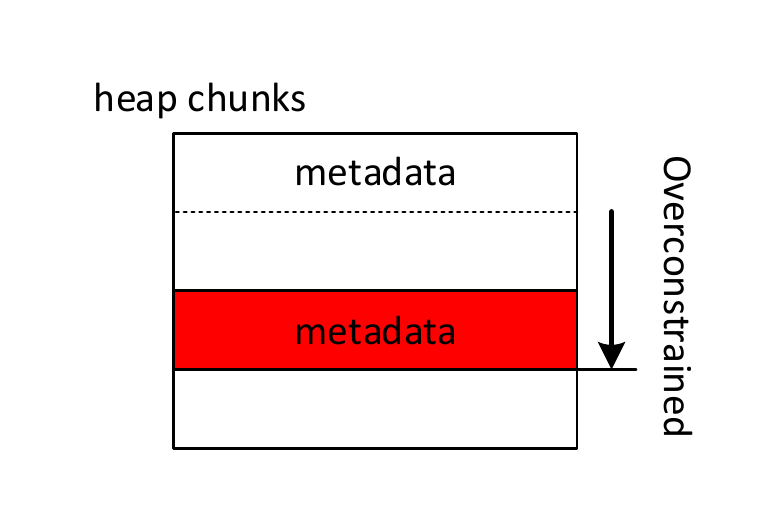}
	\caption{Heap Overflow}
	\label{fig:heapbugs}
\end{figure}

\subsection{Bug Types}

In this work, we focus on two different memory safety errors: \emph{stack buffer overflows} and \emph{heap buffer overflows}. These bug types were chosen because they are already implemented in LAVA, and are still very common in software written in unsafe languages such as C.

Figure~\ref{fig:stackbugs} shows a stack-based buffer overflow. In this type of overflow, a buffer on the stack that stores attacker-controlled data can be overflowed, causing the attacker's data to overwrite other variables on the stack, the saved frame pointer, and the return address. Because the stack layout of a function is determined at compile time, we can control what data will be overwritten when the overflow occurs, which gives us an opportunity to ensure the overflow will not be exploitable. Stack-based overflows that overwrite the frame pointer or return address will manifest as soon as the function returns, whereas those that overwrite another local variable will have more complex effects that depend on how the overwritten local variable is used by the program after the overwrite.

In a heap-based buffer overflow, shown in Figure~\ref{fig:heapbugs}, an attacker can overflow a buffer allocated on the heap, causing heap metadata (e.g., freelist pointers or size fields) stored in between the heap chunks to be overwritten. These bugs may manifest when the heap implementation interacts with the corrupted heap chunk (e.g., when the chunk is freed or a new allocation is requested). Unlike the stack, the layout of the heap is not known at compile-time; as we will explain, this means that extra care must be taken to ensure non-exploitability, and some heap bugs cannot be made non-exploitable and must be avoided.

\subsubsection{Bug Injection}

Recall from Section~\ref{sec:background} that the bugs LAVA injects consist of DUAs (program inputs that are dead, uncomplicated, and available) and attack points. LAVA creates bugs by \emph{siphoning} the input data from the DUA into a global variable and then later retrieving that data at some point in the program where we want to trigger a bug.

Each chaff bug we inject uses two DUAs. The first, which we will call the \emph{trigger DUA}, is used as a trigger; if the input data does not satisfy some condition (e.g., equality with some constant), then the bug will not be activated. This allows us to ensure that the bug does not manifest for too many inputs and thereby cause the program to fail on benign inputs. The second DUA (the \emph{attack DUA}) is used as the source buffer for the overflow. As with the original LAVA system, both DUAs should have low \emph{liveness} (number of branches influenced by the input data) and \emph{TCN} (taint compute number, a measure of how much computation has been done on the data).

Unlike LAVA, however, we have an additional constraint on how we can select DUAs. The DUAs selected must be fully initialized at the point where we wish to siphon them into global data. If we fail to do so, we may inadvertently add unintentional bugs (such as a use-after-free or a read of an uninitialized local variable) to the program; since we do not intend to inject these bugs we cannot guarantee that they are non-exploitable. We discuss our technique for avoiding uninitialized DUAs in more detail in Section~\ref{sec:uninit_dua}.

We have considerable freedom in where we place our attack points. We can create an attack point at any point along the execution trace for our seed inputs as long as the attack point is reached after the trigger DUA. This allows us to place bugs in any part of the program that can be reached by the inputs we have available; in Section~\ref{sec:eval:ss:func} we show experimentally that we can, in fact, place bugs throughout the program given an adequate set of test inputs.

\subsection{Ensuring Non-Exploitability}

When injecting an overflow, two key quantities determine exploitability: the
\emph{target} of the overflow (i.e., what data is overwritten) and the
\emph{value} that is written. Each of these can be manipulated so that
exploitation becomes impossible. With this in mind, we designed two strategies
for ensuring our bugs are non-exploitable: making the target of the overflow a
variable that is \emph{unused} by the program, and ensuring that the value used in the
overflow is \emph{constrained} so that it can only take on safe values at the site of
the overflow. In principle, each strategy can (with some care) be used with
both stack- and heap-based overflows, yielding four classes of non-exploitable
bugs, but in practice heap-based unused variable bugs cannot be guaranteed to
be non-exploitable and must be excluded. In this section we discuss our
strategies for ensuring non-exploitability in detail.

\subsubsection{Unused Variable Overwrite}

Our first strategy is to arrange the overflow so that it can only overwrite a variable that is unused by the program. Although it is difficult in general to determine statically whether a given variable is unused, as the bug injector we have a significant advantage: rather than trying to find an existing unused variable we can simply add our own variables and then overwrite them. We note that these overwrites will not cause a program crash, but they should be detectable through either static analysis or fine-grained bounds checking.

For stack-based overflows, the unused variable must also be on the stack---in other words, it must be a local variable. We rely on the compiler to guarantee the two variables (the buffer we overflow and the unused variable) are allocated adjacent to one another. In our current implementation, we simply rely on our knowledge of how our chosen compiler lays out local variables on the stack, but for a stronger guarantee we could also extend the compiler to provide annotations that allow the stack layout to be specified explicitly.

Heap-based unused variable overwrites seem plausible at first, but cannot in practice be made non-exploitable.
Ensuring the safety of an overflow requires that we be able to precisely predict what will be overflowed. However, for a heap-based unused variable overflow, we would need to guarantee that the following heap chunk is one we allocated. Because the placement of a given chunk depends on the current state of the heap, which in turn depends on the sequence of allocations and deallocations made in the program so far, we can make no such guarantee, and so these bugs might be exploitable. Hence, we exclude them from our system.

\subsubsection{Overconstrained Values}
\label{sssec:overconstrained}

Our second strategy ensures non-exploitability by using overconstrained values. Because we control the data-flow path by which the malicious input (i.e., the attack DUA) reaches the attack point, we can add constraints to the data that restrict it to a set of known-safe values. This ensures that the bug can be triggered, but only with values that we can prove do not lead to code execution.

Overconstrained values can be used to overwrite any sensitive data as long as a safe subset of values can be found. In our prototype, our bugs overwrite sensitive data such as the stored frame pointer, return address or heap metadata. The return address is used to restore program counter after function returns, so constraining the value to addresses in unmapped or non-executable ranges is safe. Similarly, if we overwrite the frame pointer saved on the stack, then it will be restored (into \verb+ebp+ on x86) when the function returns; since the restored frame pointer will be dereferenced to access local variables (e.g. \verb+mov eax, [ebp-0xC]+), values that result in unmapped addresses are once again safe.

Finally, for heap overflows we can target the heap metadata. Safe values for metadata overwrites vary depending on the allocator, so we assume that the defender can specify which heap allocator is used. Safe values here are those that will be detected by the allocator's sanity checks; for example, we can overwrite the \verb+size+ field of the metadata for a heap chunk with the value \verb+0+. When the chunk is freed, the allocator will detect that the chunk size is invalid and safely terminate the program. This type of overwrite requires some care to ensure that it is not exploitable regardless of the state of the heap, which requires careful analysis of the heap allocator implementation. We provide such an analysis in Section~\ref{sec:impl:ss:heap} for the GNU \texttt{glibc} \texttt{ptmalloc} implementation.

\subsection{Increasing Triage Difficulty}

Upon finding a bug, an attacker must triage it to determine its exploitability. In order to increase the attacker's work factor, we must therefore make our bugs difficult to triage. This requires some additional work, as a na\"ive implementation of the above strategies will result in bugs that are relatively simple to triage. Overconstrained values can be easily identified as non-exploitable as long as the attacker is able to find the constraint checking code. Unused variable overwrites are also easy to triage since our unused variable's value will not escape the current function scope; this means that an attacker only needs to look at a single function to tell whether the overwritten value could be used for an exploit.

To increase triage difficulty, we need to obfuscate the constraints used in overconstrained value bugs and create additional dataflow that propagates our unused variables' values beyond the scope of the function containing the overflow. In general, we want each bug to require reading and understanding significant amounts of additional code before and after the point where the bug is triggered.

\begin{figure}
	\centering
	\includegraphics[width=0.5\textwidth]{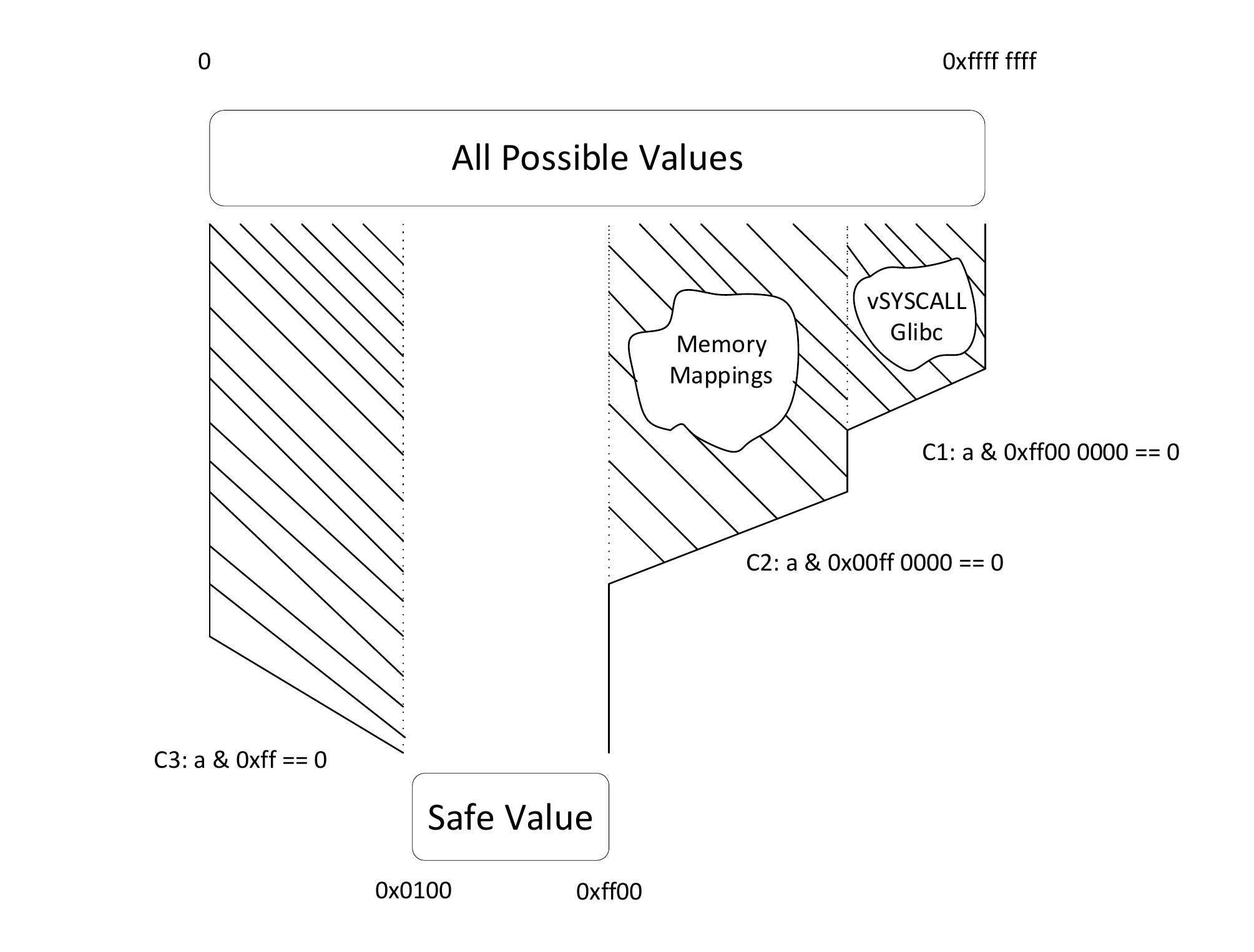}
	\caption{An overconstrained value bug. By adding constraints along the path leading to the bug, we gradually eliminate unsafe values.}
	\label{fig:ovrcondiag}
\end{figure}

\begin{figure}
	\centering
	\includegraphics[width=0.5\textwidth]{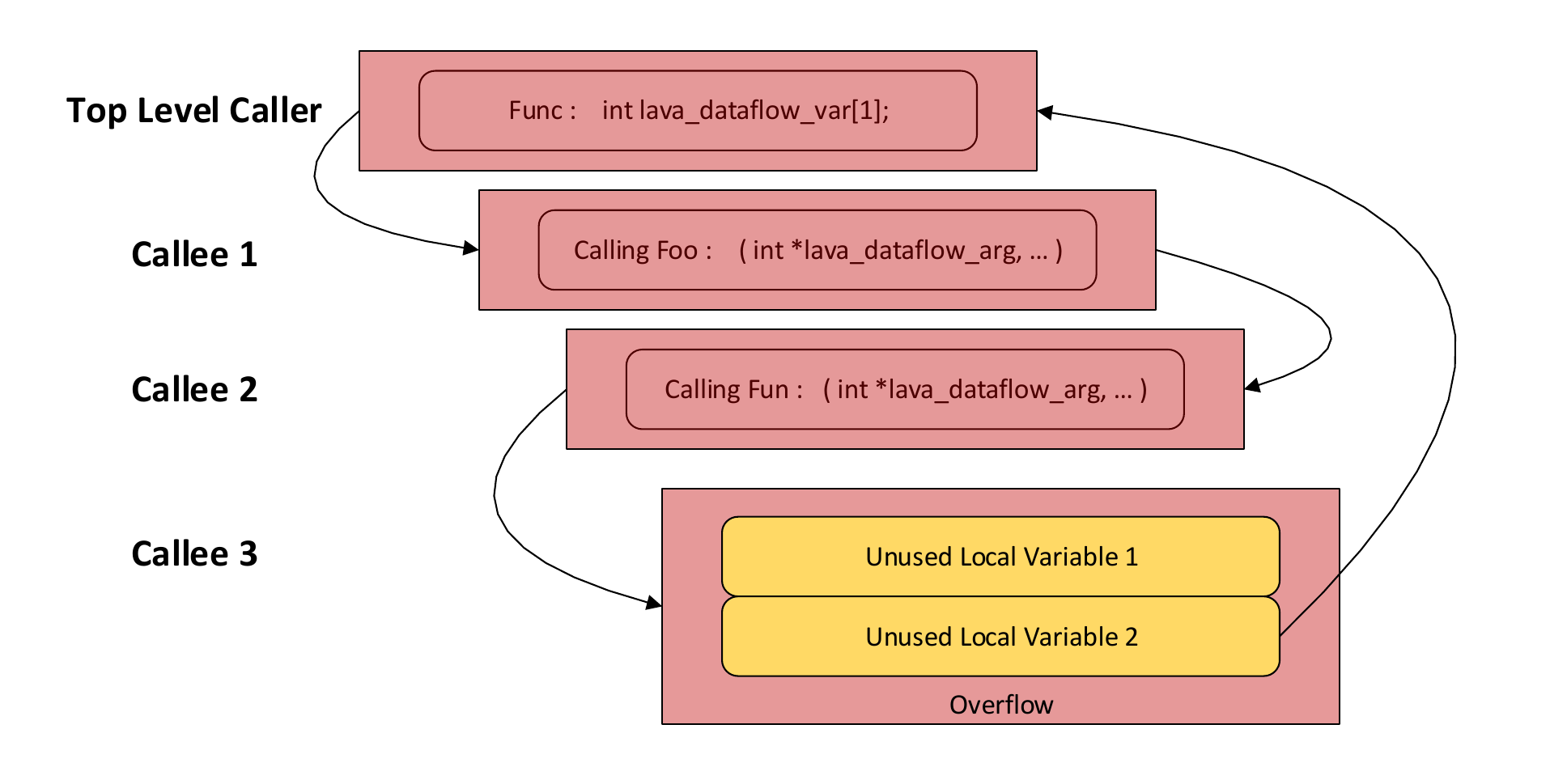}
	\caption{An unused variable bug. Data flow is added to propagate the ``unused'' values outside the initial scope, hiding the fact that they are actually unused.}
	\label{fig:unuseddiag}
\end{figure}

\subsubsection{Overconstrained values}

To obfuscate constraints, we split the constraint check into multiple parts and spread the parts throughout the execution trace that leads to the bug. Each check along the way excludes some subset of unsafe values before propagating the input onward, as shown in Figure~\ref{fig:ovrcondiag}. To check whether the bug is exploitable, an attacker has to reconstruct the exact path that would reach the bug and verify that no other path can reach the bug with more permissive constraints. Each check in isolation does not rule out all unsafe values, so unless an attacker considers all of them he cannot rule out the possibility that the bug is exploitable.

One concern with overconstrained values is the possibility that more advanced techniques, such as hybrid concolic fuzzers~\cite{Stephens:2016}, could use an SMT solver to determine that the input that triggers a crash is constrained to safe values and automatically discard it. However, we note that ruling out a bug as non-exploitable requires reasoning not just about the path leading to a crash, but reasoning about all other paths that may lead to the same crash but with fewer constraints; failing to do so would discard real, exploitable bugs as well. Additionally, we note that in principle arbitrarily complex functions could be constructed that map the original attacker-controlled input to a safe subset of values, and that the associated constraints could be constructed to make them difficult for SMT solvers to check. Such functions are a generalization of opaque predicates~\cite{Collberg:1998}, which are known to be difficult to analyze.

\subsubsection{Unused variables}

In order to make triage harder for the unused variable strategy, we must make it appear that the unused variable is, in fact, used by other parts of the program. To do so, we add data flow after the overflow takes place that carries the overwritten value to other parts of the program (but, ultimately, does not use it). As shown in Figure~\ref{fig:unuseddiag}, we add additional output arguments to the function that contains the unused variable bug and a fixed number of its parents. After the overflow, we add code to copy the overwritten value into this output argument and propagate it back up the stack of calling functions. This increases the difficulty of triage by forcing the attacker to check how the value is used in all the callers. This process could be extended by introducing dataflow from the calling functions to later callees; for our current prototype, however, we only create dataflow to the calling functions.

\section{Implementation}
\label{sec:impl}

In this section we discuss the details of our implementation. We implement our non-exploitable bug injection system on top of LAVA~\cite{Dolan-Gavitt:2016}, a previously published bug injection system that is itself based on the PANDA dynamic analysis platform~\cite{panda} and the clang compiler~\cite{clang}. Our implementation currently assumes a 32-bit x86 Linux environment, but the ideas are general and could be adapted to other operating systems and architectures with minor modifications.

\subsection{Extensions to LAVA}

Apart from the changes to LAVA to make bugs non-exploitable (which are discussed in detail below), the major extensions to LAVA involve fixes to avoid \emph{unintended} bugs. The original LAVA system can, in some cases, add accidental bugs such as use-after-frees and uninitialized variable reads. For this work, we eliminated these unindented bugs by taking care not to use any uninitialized DUAs and adding extra safety checks around pointer uses.

We also extended PANDA to support tracking taint in programs that take input from the network rather than a local file, and added monitoring of heap allocations and stack layout to PANDA's analyses.

Overall, our changes to LAVA comprise 7,935 lines added and 325 deleted across 82 files in the LAVA codebase (as somewhat crudely measured by \texttt{git diff}). Our system will be released under an open source license upon publication.

\subsection{Bug Injection}

We implemented two types of non-exploitable bugs: 1) bugs overwriting unused data, and 2) bugs overwriting sensitive data but over-constrained to non-exploitable values. We implement the unused data overwrite bugs as a stack overflow, overwriting an unused local variable, and implement both stack- and heap- based overflows with the overconstrained value strategy.

\subsubsection{Stack-Based Overflows}

For the unused variable strategy, we ensure that the overwritten variable is unused by inserting dummy local variables so that the overflow does not corrupt other local variables. To achieve this, it is important to know the order of local variables on the stack since we have to make sure the dummy variables are placed directly adjacent to the variable we are going to overflow. In our implementation, we took advantage of the fact that \verb+clang+ allocates local variables sequentially from the higher to lower stack addresses in the order of their declarations in the source. \verb+gcc+, by contrast, uses a more complicated reordering strategy~\cite{sharma2003optimal} that makes the order hard to predict. This solution is somewhat brittle since it depends on the specific compiler version used; in future work, we hope to implement a compiler extension that allows us to explicitly specify a stack layout at compile time.

In the overconstrained value strategy, overflowing the stored frame pointer and return address requires knowledge of where on the stack the target is relative to the buffer we're overflowing. We implement this by placing the buffer we will overflow at the bottom of the current stack frame; during the taint analysis phase of bug injection we also measure the exact distance between the end of this buffer and the frame pointer and return address. This distance is then used to determine the maximum amount by which we will overflow the buffer, ensuring that we cannot accidentally overwrite data in previous frames.

We observed that in some functions arguments will be first copied to the bottom of the stack and local variables will be allocated above them. Overflowing the stored frame pointer or return address would also overwrite these stored arguments. As we want to focus our overconstrained target on only the stored frame pointer and return address, we add an offset to local variable we are going to overflow when the bug is triggered. This skips over the copied arguments and overwrites precisely the target data we want to corrupt. As with the ordering of stack variables, this workaround could be avoided with minor changes to the compiler.

\subsubsection{Heap-Based Overflows}
\label{sec:impl:ss:heap}

Our current implementation uses GNU \texttt{ptmalloc} as found in Debian 9.3's \texttt{glibc} 2.24\footnote{Note that this version includes a patch by Chris Evans that adds an extra assertion to \texttt{unlink}; our analysis below depends on this assertion being present.} because it is a widely used, real-world allocator. The heap overflows we inject corrupt the metadata for heap chunk just after an allocation we make, and so our analysis of the values that are safe to use in the overflow depends on the specific allocator in use. We believe the analysis in this section is correct, but we acknowledge that the argument is fairly detailed and it is possible that we have missed a case that would allow exploitation. However, we note that stronger assurance could be gained by implementing and using our own allocator with simplified semantics that make it easier to prove our bugs are non-exploitable. We leave the design of such an allocator to future work.

In our implementation, we create a heap allocation and then overflow it, corrupting the \texttt{prev\_size} and \texttt{size} fields of the subsequent heap chunk. We set \texttt{prev\_size} to the constant \texttt{12} and then use the overconstrained value strategy to set the value of \texttt{size} to \texttt{0}. In addition, we set the second-to-last 4-byte word of our allocation to the constant value \texttt{16}.

\begin{figure}
  \centering
  \includegraphics[width=0.45\textwidth]{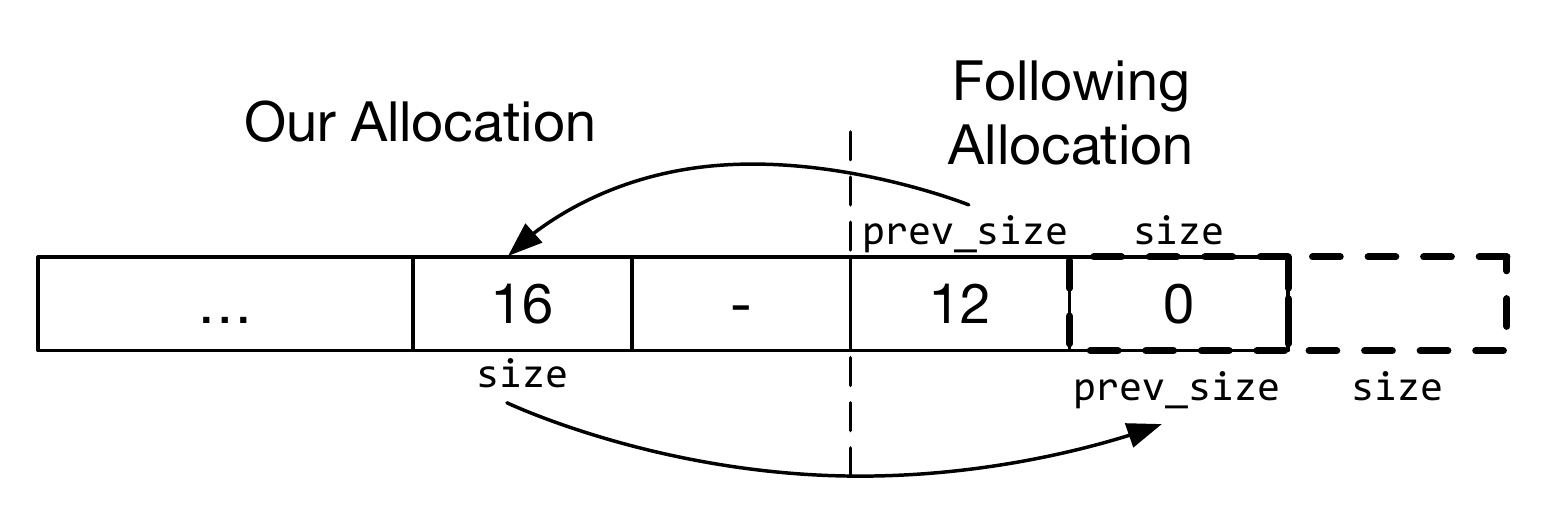}
  \caption{Layout of the heap for heap-based overflows.}
  \label{fig:heapchunk}
\end{figure}

By analyzing the implementation of \texttt{ptmalloc}, we can verify that heap overflow bugs are non-exploitable for our chosen overflow values. We split the analysis into three cases, depending on what type of heap chunk follows our allocation: an in-use chunk, the ``top'' chunk, or a freed chunk. The final case needs particular care, because other allocations may trigger \emph{heap consolidation}, where adjacent free chunks are merged together. In all cases, we need to guarantee that the allocator detects an inconsistency in the heap metadata and aborts the program, preventing the memory corruption from being exploitable.

The first two cases are relatively simple. If the next chunk is in use, the only way to trigger the bug is to free the corrupted chunk. In this case, \texttt{ptmalloc} aborts with an invalid size assertion in the free operation. If the next chunk is the top chunk, then the next memory request (i.e., \texttt{malloc}) from the top chunk will abort as intended because setting the \texttt{size} field to a small value will make the program fail in the verification of the size of the top chunk and raise an assertion error.

In the final case, the next heap chunk is a free chunk. The allocation can be in either the fast bins, unsorted bins, or the normal bins (including small bins and large bins) according to their size and the state of the heap. The corrupted heap metadata could be referenced either in a subsequent memory allocation or during chunk consolidation. For \texttt{malloc} operations, \texttt{ptmalloc} will trigger an assertion failure when the memory allocator checks that the \texttt{size} field is not less than \texttt{16}.



During chunk consolidations, \texttt{ptmalloc} will reference the heap chunk metadata to merge adjacent freed chunks to reduce heap fragmentation. To ensure an assertion failure is triggered, we make use of the fact that the \texttt{unlink} operation verifies that the \texttt{size} field of the current chunk is equal to the \texttt{prev\_size} field of the next chunk. Our heap overflow sets the \texttt{prev\_size} field of the next chunk to \texttt{12}, which points to a ``fake'' heap chunk inside our allocated buffer with a \texttt{size} field equal to \texttt{16}. The fake chunk, in turn, points to a heap chunk whose metadata partially overlaps the metadata we originally overflowed---in particular, it points to a heap chunk whose \texttt{previous\_size} field is \texttt{0} (this is depicted in Figure~\ref{fig:heapchunk}). This violates the invariant ``\texttt{chunk.next.prev\_size == size}'', so an assertion failure will be triggered as desired.

Based on our reading of the \texttt{ptmalloc} code, these cases fully account for the behavior of the allocator after we have overflowed the next chunk's metadata with our specified values. All cases result in an assertion failure, which is not exploitable, so we can say that our heap-based overflows are not exploitable with the values we used. It is probable that further analysis would find other values that could safely be used as well (allowing us greater freedom in what attacker-controlled values we can permit and improving the realism of the injected bugs), but we defer this analysis to future work.

\subsubsection{Handling Uninitialized DUAs}
\label{sec:uninit_dua}

LAVA's taint analysis operates at the binary level does not take into account function scope. As a result, the DUAs found by LAVA are not guaranteed to be initialized in all program executions. An example of this is shown in Listing~\ref{lst:uninitset}: in the execution trace observed by LAVA, \texttt{cond} is 0 and so \texttt{pointer} is initialized and detected as a DUA. However, if we then try to inject DUA siphon code after the if statement, there may be executions in which this creates an uninitialized use of \texttt{pointer}. This would add an unintended (and potentially exploitable) bug to the program.

\begin{listing}[ht]
\centering
\begin{minted}{c}
if (cond == 0) {
  Initialize(pointer);
}

lava_set(pointer);
\end{minted}
\caption{Example of an uninitialized DUA siphon. We may encounter an uninitialized use of \texttt{pointer} when \texttt{cond} is 0.}
\label{lst:uninitset}
\end{listing}

To handle this issue, we restrict the DUAs we use to those that we can guarantee have been initialized by only using DUAs once we have seen them referenced in the same scope or in an enclosing scope. This strategy guarantees that we do not add any \emph{new} uninitialized uses to the program (though if the program already contains an uninitialized use, we may add additional uses of the same uninitialized variable). In the case of Listing~\ref{lst:uninitset}, the DUA will be discarded because we cannot find a properly initialized point in the scope.

This solution guarantees that the pointer itself is safe to dereference, but in many cases the DUA identified by LAVA will in fact be an \emph{offset} from a base pointer. In particular, string buffers often contain input data that can be used for DUAs, but string pointers in C point to the first character of the string. Because we have no way of knowing how large the area pointed to by a C string pointer may be statically, we instead insert runtime checks that call \texttt{strlen} and only attempt to reference offsets less than the string's length. This results in many more available DUAs than simply discarding all character pointers with offsets, and is important for achieving amounts of chaff in programs that process text-based input such as \texttt{nginx}. Our pointer safety checks can be seen in Listing~\ref{lst:checkduanull}.

\begin{listing}[ht]
\centering
\begin{minted}{c}
if (pointer->field) {
  ...
}

temp_pointer = pointer->field;
...

// lava_set(*(pointer->field));
if (pointer->field &&
    strlen(pointer->field) > offset)
  lava_set(*(pointer->field) + offset);
\end{minted}
\caption{Example of a safe DUA siphon}
\label{lst:checkduanull}
\end{listing}

\subsection{Increasing Triage Difficulty}

\subsubsection{Faking Data Flow}

For the unused variable strategy we increase triage difficulty by adding additional uses of the unused variable after the site of the overflow. This ensures that the exploitability of the bug cannot be determined by simply observing that the corrupted value does not escape the local scope; instead, an attacker must trace the flow of data after the overflow through several other functions. Since a given function may have many call sites, this can significantly increase the amount of work required for the attacker.

Our implementation first collects the call stack to get a trace of callers at the attack point. We then filter the caller trace to eliminate library calls and the ``main'' function, since these functions are externally visible and their signatures cannot be safely modified. An extra integer local variable is added at the root of the call trace and is passed by reference to each function in the call stack by adding an extra parameter to each function between the root and the overflow site. At the overflow site, the overflowed variable is assigned to this parameter. Finally, for every modified function, we must also modify its callers so that they also use our extra parameter.

Currently, our implementation does not attempt to modify any function that is called via a function pointer, because it is difficult to statically determine all possible functions that might be assigned to such a pointer. This leads to some reduction in our ability to create fake data flow; however, in \texttt{nginx} we found only 310 indirect calls that we might miss out of 4954 total function calls, so we believe the impact is negligible.

\subsubsection{Overconstrained Value Obfuscation}

Our implementation of overconstrained value obfuscation is currently designed to successively add constraints to an attacker controlled input to eventually force its value to be zero. It does so by propagating data from the DUA to the attack point in several stages. Each stage uses a bitmask to eliminate some of the attacker controlled data and then copies the result onward. Because our bugs are created starting from a concrete input, we know the full execution path between the DUA and the attack point, and code to selectively propagate the data can be inserted along this path.

An example of the code we insert can be found in Listing~\ref{lst:ovrconcode}; the first stage clears the upper 16 bits of a 32-bit value, and the second stage clears the lower 16 bits, leaving a constant value of zero. However, to determine exploitability, an attacker must find and understand these statements, realize that they produce a constant value, and check that there are no other paths through the binary that reach the bug while retaining the attacker-controlled data.

\begin{listing}
\begin{minted}{c}

// Initial Store of DUA
setConstraint(DUA_2);

// Update
// Update first 16 bits
v1 = DUA_2 & 0xffff;

// Update last 16 bits
v2 = v1 & 0xffff0000;

// Final DUA check and trigger
if (DUA_1 == trigger)
  triggerBug(v2);
\end{minted}
\caption{Overconstrained DUA propagation. In this example, we want to constrain DUA\_2 to 0 when the bug is triggered. v1 and v2 are global variables initialized to be null. So whatever the order of ``Update'' code pieces being reached (even if some of them are skipped in different execution paths), we can still be sure that the final data used to trigger the bug has been properly constrained to 0.}
\label{lst:ovrconcode}
\end{listing}

\section{Evaluation}
\label{sec:eval}

Our evaluation focuses on answering three key questions about chaff bugs:
\begin{itemize}
  \item Can we add a large number of chaff bugs distributed throughout a given program, and will it still work correctly once we have done so? (Section~\ref{sec:eval:ss:func})
  \item What is the performance impact of adding chaff bugs? (Section~\ref{sec:eval:ss:perf})
  \item How successful are current bug-finding tools at finding chaff bugs, and do automated triage tools think they're exploitable? (Section~\ref{sec:perf:ss:triage})
\end{itemize}

\subsection{Functionality}
\label{sec:eval:ss:func}

We injected chaff bugs into three target programs: \texttt{nginx} 1.13.1, \texttt{file} 5.30, and \texttt{libFLAC} 1.3.2. The bugs are selected randomly among the three types of bugs our prototype creates: overconstrained stack (OC-Stack), overconstrained heap (OC-Heap), and unused stack (Unused-Stack). For each bug, we verify that it was successfully injected by running its triggering input and observing a crash. Because our unused variable bugs do not trigger crashes, we modified them to additionally raise an exception by dividing by zero.

We generated two sets of inputs for each program: clean inputs, which come from the corpus we used for bug generation (e.g. the official \texttt{nginx} test suite, \texttt{libFLAC} testing examples, etc.), and triggering inputs, which are modified versions of the clean inputs that will trigger one of our bugs. For all clean inputs, we verified that the buggy programs are able to process them with the expected output. For all selected triggering inputs, we checked that the programs crash as intended. Because there are so many candidate bugs, we were only able to test a subset. Table~\ref{tab:crasheval} shows the number and type of bugs we verified.

Two main factors account for the fact that some bugs cannot be successfully injected. First, a candidate bug may be eliminated when we ensure that its DUA does not reference an uninitialized variable (Section~\ref{sec:uninit_dua}); this also explains why there are generally fewer overconstrained bugs---overconstrained bugs use two DUAs (one for the trigger and one for the overconstrained value), so there are twice as many opportunities for them to be eliminated. Second, overconstrained bugs targeting the frame pointer may not be triggerable because the calling function may not reference any local variable before it returns, at which point its (non-corrupted) frame pointer will be restored, avoiding a crash.

\begin{table}
	\centering
	\caption{Validated Bugs}
	\label{tab:crasheval}
  \begin{tabular}{lccc}
    \hline
    & \texttt{nginx} & \texttt{flac} & \texttt{file} \\
    \hline
    Attempted & 27400 & 176500 & 132091 \\
    OC-Stack & 192 & 554 & 23039 \\
    OC-Heap & 853 & 570 & 24632 \\
    Unused-Stack & 5178 & 2045 & 45547 \\
    \hline
	\end{tabular}
\end{table}

We next measured our ability to add bugs to all parts of the target program. If all of our injectable bugs are concentrated in some small part of the program, an attacker might be able to simply ignore any bugs found in that area and find preexisting, exploitable bugs elsewhere.\footnote{In fact, it would be sufficient to be able to place chaff bugs wherever naturally occurring bugs exist in the program, but since we don't know the distribution of naturally occurring bugs for a given program we have to conservatively assume they could be anywhere.}
  
First, we measured the baseline coverage of our test corpus for each program; because LAVA (and by extension our system) relies on dynamic analysis, it cannot add bugs to parts of the program that are not covered by the inputs in the test corpus. We computed the function and file coverage using gcov 4.9.2 (lcov 1.1). All programs were configured with optional modules disabled and compiled with clang 3.6.2.
  
The original inputs' coverage is given in Table~\ref{tab:origincov}. The attack point (ATP) coverage columns indicate potential sites where injected bugs may be triggered in our system. ATP function coverage is relatively low; however, this is partially because attack points can only occur after attacker-controlled data is introduced to the program. If we exclude functions that are executed before the first byte of input is read, the function coverage rises to 56.0\% for \texttt{nginx}, 80.3\% for \texttt{file}, and 65.7\% for \texttt{flac}. Higher coverage may be achievable by changing configuration settings for the programs, trying different command line parameters, etc., but the problem of finding high-coverage tests for a program is both difficult and orthogonal to our current research.

\begin{table*}
\centering
\caption{Coverage of Original Inputs}
\label{tab:origincov}
  \begin{tabular}{lccc}
    \hline
	& Functions & ATP file coverage & ATP function coverage \\
	\hline
	\texttt{nginx} & 52.8\% (712/1349) & 45.6\% (47/103) & 26.4\% (188/712) \\
	\texttt{file} & 54.8\% (142/259) & 63.2\% (12/19) & 40.1\% (57/142) \\
	\texttt{libFLAC} & 19.3\% (145/752) & 30.2\% (13/43) & 46.2\% (67/145) \\
	\hline
\end{tabular}
\end{table*}

Finally, we ran an experiment to estimate the number of attack points that were actually usable in practice. We randomly selected 100 unique attack points in each program. Each attack point may be associated with many bugs (i.e., many different DUAs). We then attempt to inject bugs using each attack point. If a given bug cannot be triggered at that attack point, we try again with a different bug, until all bugs have been exhausted.

\begin{figure}
	\centering
	\includegraphics[width=0.5\textwidth]{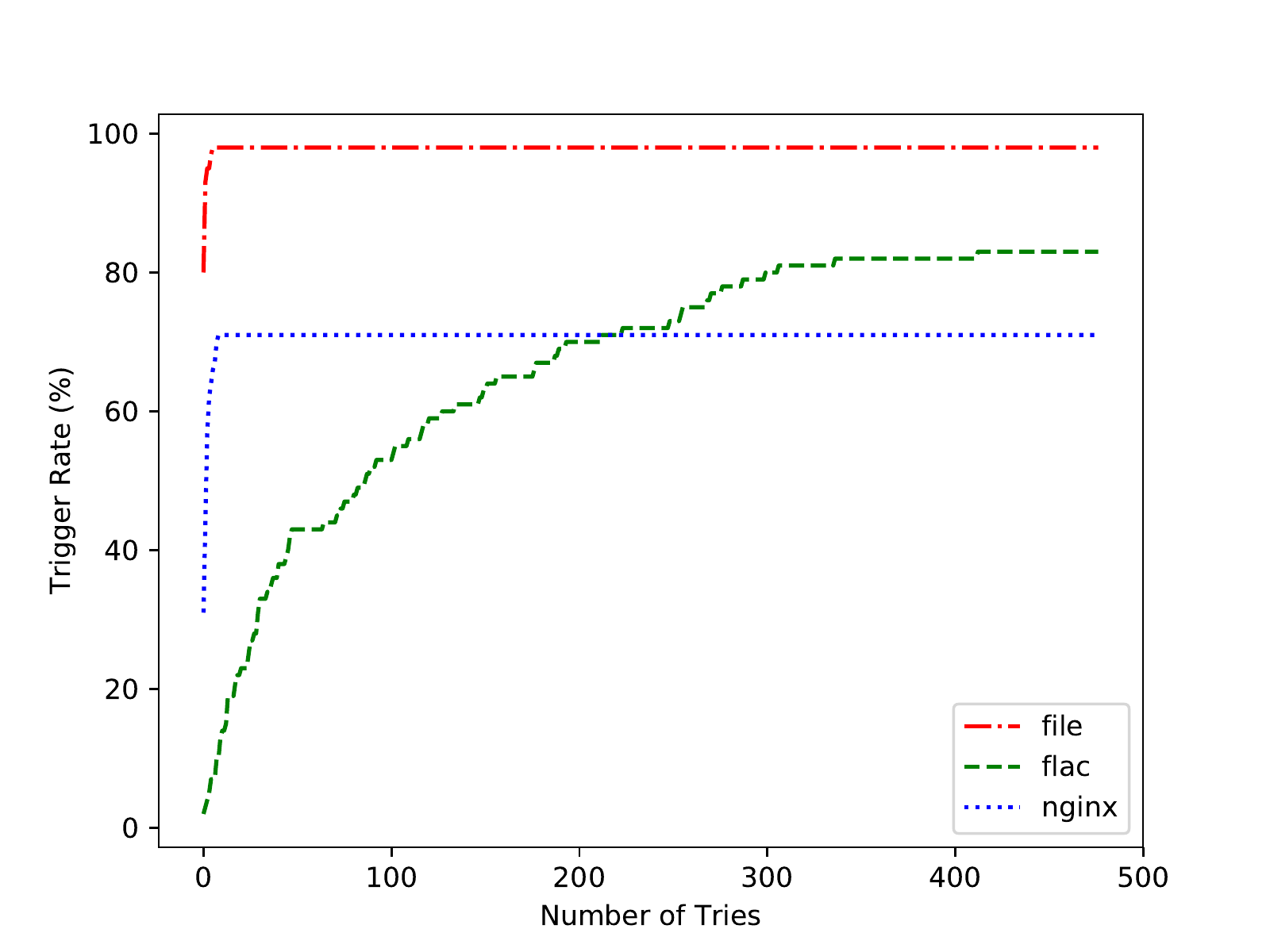}
	\caption{Success rate at triggering bugs for each ATP}
	\label{fig:coveragefig}
\end{figure}

Figure~\ref{fig:coveragefig} shows the success rate after $k$ tries. We find that \texttt{nginx} and \texttt{file} have a relatively higher trigger rate in the beginning. With multiple tries of bug injection, \texttt{file} achieved a 98\% success rate within 6 tries, while \texttt{nginx} ultimately reaches 71\%. The differences here seem largely due to programming style and the type of input processed by each program. \texttt{nginx}, for example, makes extensive use of pointer DUAs to parse string data so many potential bugs are eliminated when we check for uninitialized data usage. \texttt{libFLAC} takes longer to reach a high success rate (perhaps because of the large number of potential bugs available), but reaches 83\% eventually.

\subsection{Performance}
\label{sec:eval:ss:perf}

We measured the performance overhead of chaff bugs by benchmarking our buggy version of \texttt{nginx} with different numbers of bugs added. We ran Weighttp~\cite{apachebench} against \texttt{nginx} 1.13.1 injected with 10, 50, 200, 1000, and 2000 randomly selected bugs. The \texttt{nginx} binary is compiled with LLVM 3.6.2. Our testbed is a 24-core Intel\textregistered~Xeon\textregistered~X5690 CPU at 3.47GHz with 189 GB of RAM running Debian 9.3.

\begin{figure}
	\centering
	\begin{subfigure}{.5\textwidth}
	\centering
	\includegraphics[width=0.8\textwidth]{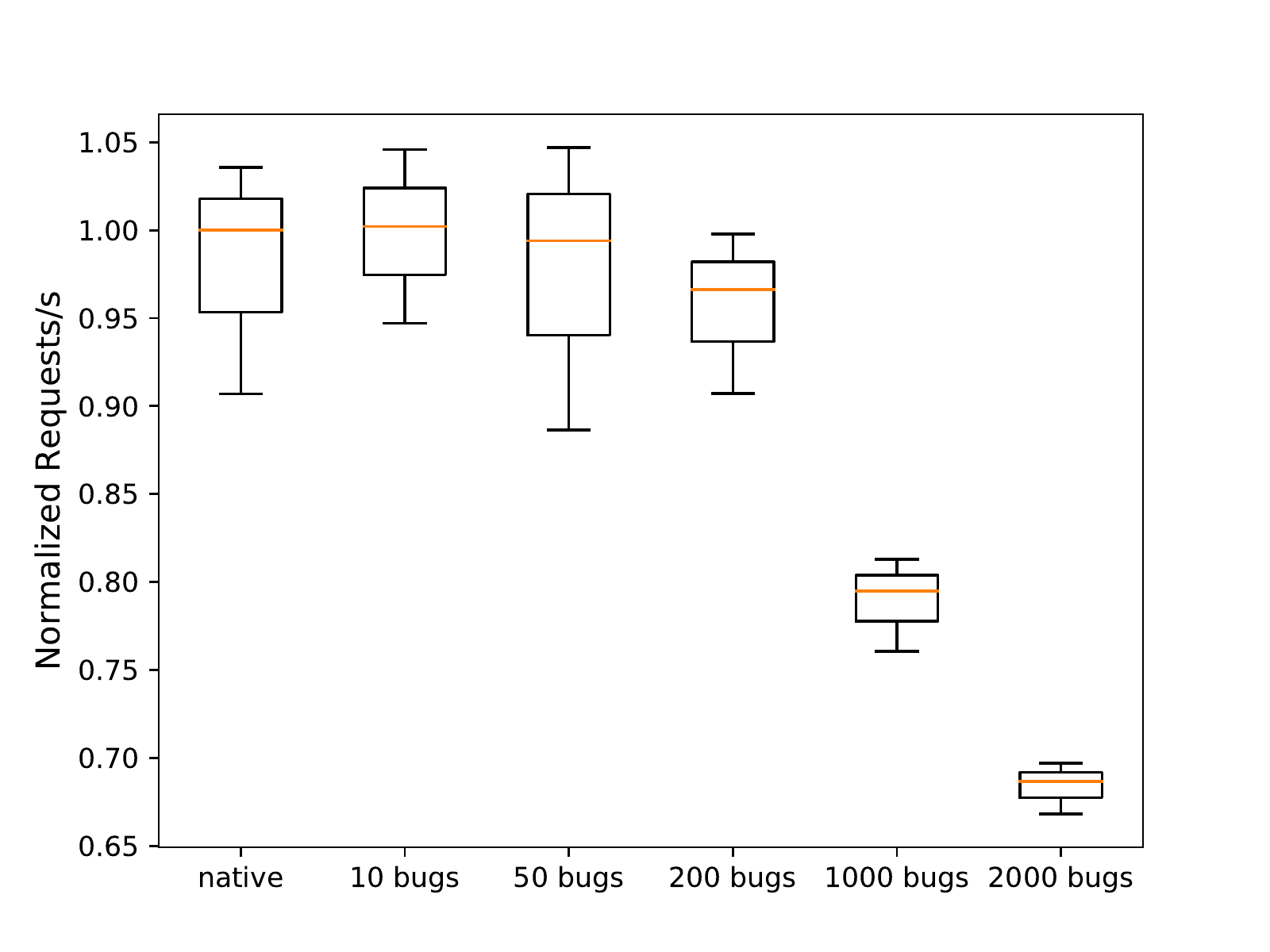}
	\caption{1 worker}
	\end{subfigure}
	\begin{subfigure}{.5\textwidth}
		\centering
		\includegraphics[width=0.8\textwidth]{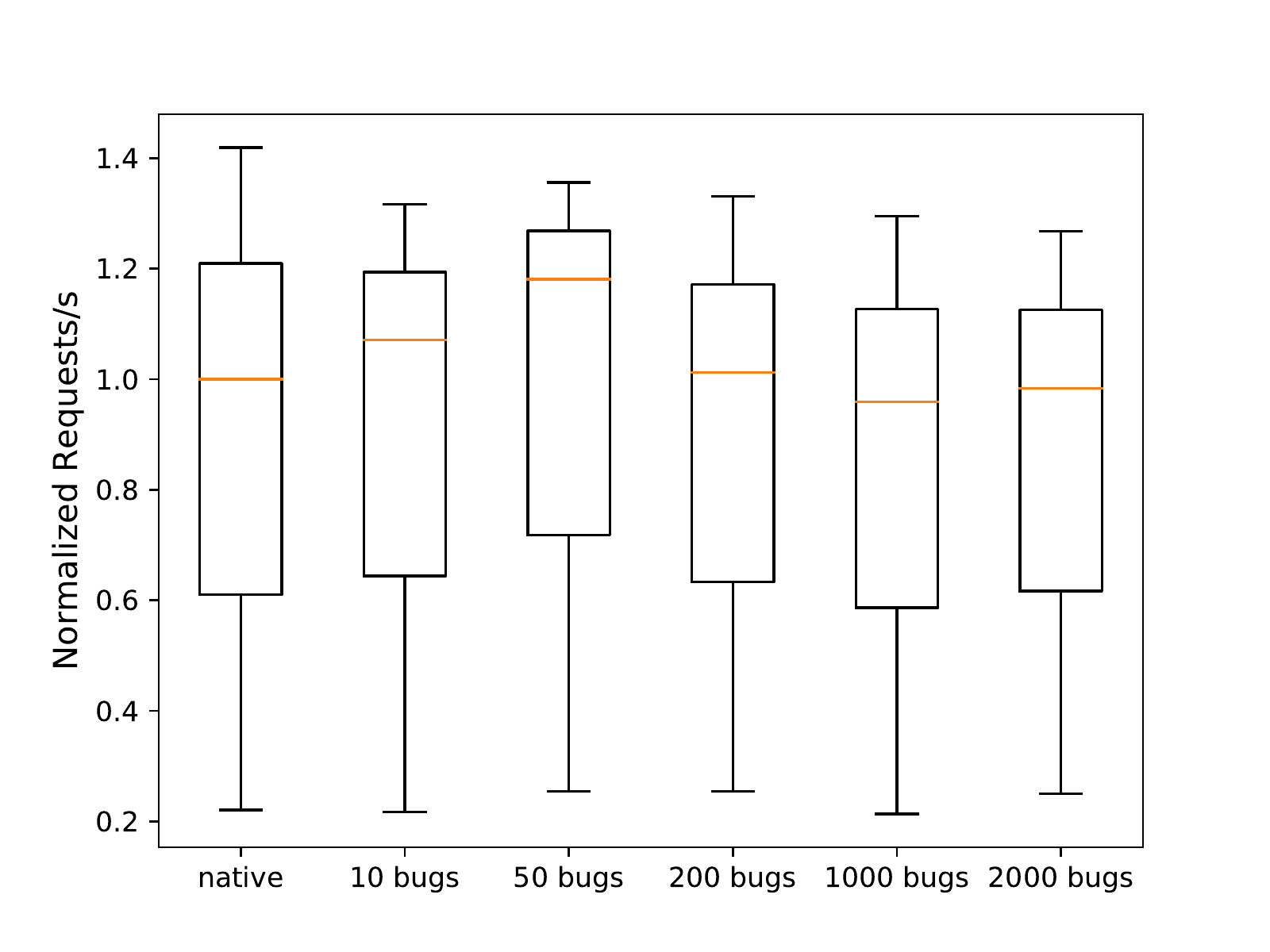}
		\caption{24 worker}
	\end{subfigure}
	\caption{Performance Evaluation for \texttt{nginx} 1.13.1}
	\label{fig:perfngx}
\end{figure}

Figure~\ref{fig:perfngx} shows the average requests processed per second by \texttt{nginx} per second with one and 24 concurrent workers. With more bugs injected, we observed higher overhead, mainly due to the added cost of checking whether inputs match our bugs' trigger conditions. For 24 workers, we observed lower overhead compared to a single worker. We speculate that with more workers available, other performance bottlenecks (such as disk and network I/O) dominate.

Profiling \texttt{nginx} with different numbers of bugs injected, we identified some injected code accounting for more performance overhead than the others---they were hit unusually often in the run. We identified three functions, for memory allocations and string formatting, in \texttt{nginx} that are extremely ``hot'' and not heavily used in bug injection; excluding these from bug injection resulted in around 15\% performance gain in the final result. Further optimizing with LLVM's Profile-Guided Optimization (PGO)~\cite{clang} would also give us an extra 20\% boost in performance.

\subsection{Bug Detection and Triage}
\label{sec:perf:ss:triage}

To get an estimate of the difficulty of finding and triaging our bugs, we used off-the-shelf tools to test our buggy binaries and triage any crashes found. We used the American Fuzzy Lop (AFL) fuzzer~\cite{afl} to find crashing test cases and then used the gdb \texttt{exploitable}~\cite{gdbexploitable} plugin to evaluate their exploitability.

Our fuzzing targets were each injected with 1000 randomly selected overconstrained chaff bugs (unused variable bugs were omitted because they do not cause crashes and hence cannot be found by fuzzing without additional instrumentation). For \texttt{file} and \texttt{flac}, we injected 500 heap-based bugs and 500 stack-based bugs; however, for \texttt{nginx} we were only able to successfully inject a total of 54 stack-based bugs, so we increased the number of heap-based bugs to 810 bugs. For each program, we generated a fuzzing dictionary from the strings and integer constants found in the target binary, and seeded AFL with a held-out input file that was not used during the bug injection step. We fuzzed each target with 11 concurrent AFL processes in parallel for 24 hours (wall clock time) and measured the number of unique crashes as determined both by AFL's built-in crash bucketing and our own ground truth knowledge of each bug.

Table~\ref{tab:fuzzing} shows the fuzzing results. For \texttt{libFLAC}, we found 1275 crashes that AFL considered unique---more crashes than there are injected bugs, indicating that some of our bugs were mistakenly counted multiple times by AFL. This is likely a consequence of the heap-based bugs we injected: we only corrupt the heap metadata when triggering the bug, so the actual crash occurs later when the program allocates or frees memory. In \texttt{nginx}, AFL did not find any of our stack-based bugs because there were fewer to find and its detection rate was relatively low overall.

\begin{table*}
	\centering
	\caption{Fuzzing Results}
	\label{tab:fuzzing}
	\begin{tabular}{lcccc}
		\hline
		& Unique Crashes & Unique Bugs & Heap Bugs & Stack Bugs \\
		\hline
		\texttt{nginx} & 55 & 55 & 55 & 0 \\
		\texttt{file} & 595 & 350 & 305 & 45 \\
		\texttt{flac} & 1275 & 315 & 155 & 160 \\
		\hline
	\end{tabular}
\end{table*}

Finally, we tested whether a popular open-source triage tool, gdb \texttt{exploitable}, considered our bugs exploitable. We used the same programs and bugs as for the fuzzing experiment, but used our ground truth triggering inputs to create crashes for \texttt{exploitable} to analyze. Table~\ref{tab:exploitable} shows the number of validated bugs and the category assigned by \texttt{exploitable}. In our experiments, all of our chaff bugs were considered EXPLOITABLE or PROBABLY\_EXPLOITABLE.

All heap-based overflows were classified as EXPLOITABLE no matter what heap metadata we corrupted in the injected heap-based bugs. Stack-based bugs that overwrote the stored frame pointer were also classified as EXPLOITABLE because the access violation happens at the destination operand of an instruction (e.g. \texttt{mov [ebp-0x10], eax} where \texttt{ebp} is NULL). For stack-based bugs, the return address overwrites were reported as PROBABLY\_EXPLOITABLE because a segmentation fault occurred on a program counter near NULL (recall that our prototype only uses \texttt{0} for its overconstrained values). A more refined overconstrained value strategy that sets the return address to non-executable or unmapped memory would likely be considered EXPLOITABLE.

\begin{table*}
\centering
\caption{Triage Tool Results}
\label{tab:exploitable}
\begin{tabular}{lcccc}

	\hline
  & Heap Bugs & Stack Bugs & EXPLOITABLE & PROBABLY\_EXPLOITABLE \\
  \hline
\texttt{nginx} & 810 & 54 & 856 & 8 \\
\texttt{file} & 500 & 500 & 500 & 500 \\
\texttt{flac} & 500 & 500 & 548 & 452 \\
  \hline
\end{tabular}
\end{table*}

In the future, better triage tools may be more successful at determining which bugs are truly exploitable. However, we note that as a design choice, triage tools tend to be weighted in favor of conservatively assuming that a bug may be exploitable (so that bugs are not missed), which proves to be an advantage in our case.

\section{Limitations \& Future Work}
\label{sec:future}

The primary limitation of our current work is that we have not yet attempted to make our bugs indistinguishable from real bugs. This means that they currently contain many artifacts that attackers could use to identify and ignore them. In future work, we hope to investigate techniques for making our bugs blend in with the surrounding code, and change their triggering conditions to something more natural than the current test against a magic value.

The diversity of bugs injected by our prototype is also somewhat limited, which could allow an attacker to identify patterns in the bugs we produce and exclude those that match the pattern (for example, they could decide to ignore heap overflows where the metadata header matches the constants used by our overflow). We believe this can be alleviated by introducing more types of bugs (such as use-after-free, TOCTOU, integer overflows, etc.) and by expanding the range of safe values in our overconstrained value strategy.

Our current bug injection requires source code, but in many cases source code for legacy systems is unavailable. It may therefore be useful to investigate whether chaff bugs can be injected at the \emph{binary} level as well. Although this is likely to be more difficult from a program analysis point of view, from the point of view of ensuring non-exploitability it may actually be easier. Exploitability depends heavily on low-level details such stack layout, alignment, etc., and these features are more readily visible at the assembly level~\cite{Balakrishnan:2010}.

Finally, when investigating automated crash triaging tools, we found surprisingly few public tools apart from Microsoft's \texttt{!exploitable} and its gdb counterpart, \texttt{exploitable}. In terms of academic study, the only work available seems to be a masters thesis by Frederick Ulrich~\cite{Ulrich:2017}. We hope that chaff bugs will help draw more academic attention to the problem of exploit triage.

\section{Related Work}
\label{sec:relwork}

There is a large body of work on mitigations for exploitation of bugs in software, including Address Space Layout Randomization (ASLR), Data Execution Prevention (DEP), stack canaries, Control Flow Integrity~\cite{cfi}, and Code Pointer Integrity~\cite{Kuznetsov:2014}. Our work represents a novel defense in software security that focuses on wasting attacker effort during the bug finding and exploit development process. It can be combined with these existing mitigations, but our non-exploitability analysis would need to be adapted to take this modified execution environment into account.

Our work also fits more broadly into the category of security through deception. Almeshekah et al.~\cite{Almeshekah} give an overview of different types of deception in security, including honeypots, honeyfiles, and honeytokens. Juels and Rivest~\cite{Juels:2013} describe the use of honeywords---fake password hashes associated with an account---to detect password cracking attempts. Araujo et al.~\cite{Araujo:2014} describe honeypatches, which patch software but transparently redirect exploit attempts to an unpatched instance in order to deceive attackers into thinking their attack has succeeded. Finally, one line of work on deception has focused on generating believable \emph{decoy data}~\cite{Voris:2013}, including decoy documents~\cite{Voris:2015} and source code~\cite{park2012software}; the goal of this work is to provide enticing but fake documents to attackers and then detect their exfiltration.

These sorts of honeypots are mainly focused on detecting and analyzing attacker actions. However, some research has also attempted to waste attacker resources upon detection. Liston first proposed the idea of a network tarpit and implemented it in the LaBrea software~\cite{labrea}. Tarpits deliberately delay network traffic responses to slow down attackers' network reconnaissance. A similar countermeasure is the ``endless file,''~\cite{endlessfile} a technique in which a large, sparse file is created on a remote system, which causes attackers to download large amounts of data while using little space on disk.

Closest to our own work is research into \emph{anti-fuzzing}, in which modifications are made to software to make it harder to find bugs through fuzzing. Miller~\cite{cmantifuzzing} described a technique for detecting that a program was being tested with a fuzzer and then triggering unique, non-exploitable crashes dynamically. These techniques were further developed by Whitehouse~\cite{nccantifuzzing}, who proposed a suite of defensive countermeasures to take when fuzzing is detected, including throwing fake errors, degrading performance, masking legitimate crashes with an exception handler. Most recently, Edholm and G\"oransson~\cite{Edholm2016} developed and evaluated a number of fuzzing detection and evasion techniques on a subset of the DARPA Cyber Grand Challenge~\cite{CGC} dataset. Our work differs in that we inject real (but non-exploitable) bugs, which can be found through any bug-finding technique, not just fuzzing. In addition, our technique cannot be defeated by finding and disabling the anti-fuzzing logic of the program.


\section{Conclusion}
\label{sec:conclusion}

In this paper, we have presented a novel approach to software security that \emph{adds} rather than \emph{removes} bugs in order to drown attackers in a sea of enticing-looking but ultimately non-exploitable bugs. Our prototype, which is already capable of creating several kinds of non-exploitable bug and injecting them in the thousands into large, real-world software, represents a new type of deceptive defense that wastes skilled attackers' most valuable resource: time. We believe that with further research, chaff bugs can be a valuable layer of defense that provides deterrence rather than simply mitigation.


\bibliographystyle{IEEEtran}
\bibliography{IEEEabrv,reference}

\end{document}